\documentclass[10pt]{article}

\usepackage[latin1]{inputenc}
\usepackage[T1]{fontenc}

\usepackage{a4wide}
\usepackage{amssymb,amsmath,fancyheadings,theorem}
\usepackage{fancybox}
\usepackage{shadow}
\usepackage{array}   
\usepackage{latexsym}
\usepackage{graphicx}
\usepackage{color}

\newtheorem{theorem}{Theorem}
\newtheorem{lemma}{Lemma}
\newtheorem{defi}{Definition}

\newtheorem{corollary}{Corollary}
\newtheorem{example}{Example}
\newtheorem{remark}{Remark}
\newtheorem{prop}{Proposition}

\newcommand{\F}[1]{{GF(#1)}}

\newcommand{\Rang}{\mbox{Rk}}

\newcommand{\Ln}[1]{(#1_1,\ldots, #1_n)}
\renewcommand{\Vec}[1]{\mathbf{#1}}

\begin{document}

\title{Properties of codes in rank metric}
\date{}
\author{P. Loidreau\footnote{Pierre.Loidreau@ensta.fr} \\
  Ecole Nationale Supérieure de Techniques Avancées (ENSTA)}

\maketitle

\begin{abstract}
We study properties of rank metric and codes in rank metric over finite fields. 
We  show that in rank metric perfect codes do not exist. 
We derive an existence bound that is the equivalent of the 
Gilbert--Varshamov bound in Hamming metric. We study the asymptotic
behavior of the minimum rank distance of codes satisfying GV. 
We derive the probability distribution of minimum rank distance for
random and random $\F{q}$-linear codes. We give an asymptotic
equivalent of their average minimum rank distance and show that random
$\F{q}$-linear codes are on GV bound for rank metric.
 We show that the covering density of optimum codes whose codewords
 can be seen  as square matrices is lower bounded by a function
 depending only on the error-correcting capability of the codes. 
We show that there are  quasi-perfect codes in rank metric over fields
of characteristic $2$. 
\end{abstract}

\section{Introduction}

Rank metric over finite fields used for purposes of correcting rank
errors in information transmission  was 
first introduced by E.~M.~Gabidulin in 1985, \cite{Gabidulin:1985}. In this metric codes of length $n$ 
can either  be viewed as sets of vectors over some extension field $\F{q^m}$ or equivalently 
as $m \times n$ matrices over  some base field $\F{q}$. The rank of the 
codewords corresponds to the usual matricial rank over $\F{q}$ for the matricial representation. 
This metric is suitable for correcting errors which can occur as matrices of upper-bounded rank.   
In the same seminal paper, E.~M.~Gabidulin constructed an optimal (for the metric) 
family of linear codes using properties of the Frobenius automorphism
of the extension. They can be considered as the equivalent for rank
metric of Reed-Solomon  codes. They have polynomial time decoding
algorithms based either on some kind of extended Euclidian algorithm
for the family of linearized polynomials.
More recently, in 1991, R.~M.~Roth  designed  a family of matricial 
codes over $\F{2}$ correcting crisscross errors up to a given
rank. These codes  are suitable for correcting errors that occur on
rows or columns of matrices 
like for example on tape storage. In fact they form the matricial
counterpart of the  family of optimal codes published  by E.~M.~Gabidulin \cite{Roth:1991}.
Other families of codes with polynomial-time decoding algorithm have
been designed by considering the projection of these codes on
subfields or on subspaces, or even by adding some random matrices to
the structure of direct product of the codes (RRC codes for example), 
see \cite{OGHA:2003,Gabidulin/Loidreau:2005}.

Apart from being suitable for correcting crisscross errors as they
might occur on tape storage, rank metric was involved  in the design of
coding-theory based public-key cryptosystems with reduced key size. The main reason was
that  the decoding complexity of random linear codes in rank metric is much higher than
in Hamming metric for the same set of parameters (compare the
complexity of algorithms described in 
\cite{Chabaud/Stern:1996,Ourivski/Johannson:2002}
to the complexity of the algorithm given in
\cite{Canteaut/Chabaud:1998} for decoding in Hamming metric). 
Therefore the size of the public-key
can be significantly reduced by keeping the same security level against decoding attacks. 
This consideration lead to the design of several public-key cryptosystems
relying on this metric ever since, see for example  
\cite{Gabidulin/Paramonov/Tretjakov:1991,Gabidulin/Ourivski:2001,OGHA:2003,Berger/Loidreau:2004,Berger/Loidreau:2005,Faure/Loidreau:2005}.

Very recently rank metric over finite fields found applications in the field of
space-time coding where the concept  of {\em diversity} states  the minimum rank of 
differences of matrices composing a code over the complex
field. Indeed, it has been shown how to
construct codes with a given diversity $d$ from codes over $\F{2}$
with  minimum rank distance  $d$, see for example \cite{Lu/Kumar:2005}. 
Part of this work was further extended in \cite{Hammons:2006}.

All these reasons and also the dispersion of results concerning rank metric
 motivated our interest in the  study of its general  properties over finite fields, 
as well as  properties of codes in this metric. Therefore, this paper does not only present 
new results but  collects and present results that have  already been
 published or are widely known about rank metric over finite fields 
and about codes in the metric.

In a first part we recall some basic facts about the metric and some
of its properties. We also give  bounds on the volumes of spheres and balls
of the metric. This result is used in the proofs further in the paper.
Concerning the codes we give the definition of minimum rank distance for non-linear and for
linear codes. We show that the operation of transposing the code is a rank preserving 
isomorphism. This enables us to deal only with the case where the length of the code is less than
the extension degree of the code. Is it not the case ?  Just take
the transposed code.

Then, on  the same model as in Hamming metric, we define the classical {\em sphere-packing}
and {\em Singleton} bound. Thanks to  the sphere-packing inequality and to the
approximations on the volume of balls,  we show that no {\em perfect codes} 
exist in rank metric.

We prove a Gilbert-Varshamov
like bound on the existence of code with given parameters, and we derive
an asymptotic equivalent of the minimum rank distance of a code
satisfying said to be on GV.

Concerning random codes, 
we  evaluate their average minimum rank distance and show that,
in well-chosen cases, the probability that the effective minimum distance
deviates from the average case tends to 0 asymptotically. More precisely, we
show that random $\F{q}$-linear codes are asymptotically on GV.

In the last part, we study codes that satisfy the
Singleton inequality. They are called MRD-codes for {\em Maximum Rank
  Distance} codes. After recalling the formula given  by Gabidulin 
on the rank distribution of linear MRD-codes, we present some simulations
showing that rank distribution of random  codes and of MRD-codes is
very similar.  In addition, we  prove that the density of {\em
  correctable} errors for MRD-codes corresponding to codes formed with
square matrices 
is lower bounded by a function depending only on the error-correcting
capability   of the code. In the special case where one considers
fields of characteristic $2$,  we show that even if there  are no
perfect codes in rank metric, {\em quasi-perfect} codes exists, that
is there exists a family of codes which are  asymptotically perfect.

\section{Properties of rank metric}

In the rest of the paper the code alphabet will  be  the finite
field  $\F{q^m}$ with $q^m$ elements where $q$ is the power of some
prime. Let  $\Vec{b} =
\Ln{\beta}$ be a  basis of $\F{q^m}$ over $\F{q}$. 
The integer $n$ will denote  the length of the code. Thus vectors of
the ambient space $\F{q^m}^n$   will be indifferently considered
as vectors with components in  $\F{q^m}$ or as $m\times n$ $q$-ary
matrices obtained by projecting the elements of $\F{q^m}$ on $\F{q}$
with respect to the basis $\Vec{b}$.

Rank norm of a vector from   $\F{q^m}^n$ is defined by 
\begin{defi}[\cite{Gabidulin:1985}]\label{Defi:RangVecteur} \hfill 

Let  $\Vec{x} = \Ln{x} \in \F{q^m}^n$. 
The rank of  $\Vec{x}$ on $\F{q}$, is the rank of matrix 
\[
\Vec{X} = \left( \begin{array}{ccc}
x_{11} & \cdots & x_{1n} \\
\vdots & \ddots & \vdots \\
x_{m1} & \cdots & x_{mn}
\end{array}
\right),
\]
 where $x_j =  \sum_{i=1}^n{x_{ij} \beta_i}$. 
It is denoted by  $\Rang(\Vec{x}|\F{q})$, or more simply by $\Rang(\Vec{x})$ 
whenever there is no ambiguity.
\end{defi}
Rank metric is the metric over  $\F{q^m}^n$ induced by the rank norm.
Given a vector $\Vec{x} \in \F{q^m}^n$   spheres and balls in rank  metric have the following expression:
\begin{itemize}
\item Sphere  of radius $t \ge 0$ centered on $\Vec{x}$:
  $\mathcal{S}(\Vec{x},t) \stackrel{def}{=} \left\{ \Vec{y} \in \F{q^m}^n ~|~ \Rang(\Vec{y - x}) = t \right\}$.
\item Ball of radius $t \ge 0$ centered on $\Vec{x}$:  $\mathcal{B}(\Vec{x},t) \stackrel{def}{=} \cup_{i=0}^t \mathcal{S}(\Vec{x},i)$.
\end{itemize} 
Since rank metric is invariant by translation of vectors, the volumes of spheres and
balls do not depend on the chosen center. Therefore to simplify
notations, we define
\begin{itemize}
  \item $\mathcal{S}_t \stackrel{def}{=}$ volume of sphere of radius
    $t$ in $\F{q^m}$. It is equal to the number of $m \times n$
    $q$-ary matrices of rank $ = t$. If $t=0$ then $\mathcal{S}_0
    =1$ and for $t=1,\ldots,\min(n,m)$ it is equal (see for example \cite{Barg:1998}) to 
    \begin{equation}\label{Eq:VolumeSphere}
       \mathcal{S}_t  =  \prod_{j=0}^{t-1}{\frac{(q^n- q^j)(q^m - q^j)}{q^t - q^j}}.
    \end{equation}
 
  \item $\mathcal{B}_t \stackrel{def}{=}$ volume of ball of radius $t$
    in $\F{q^m}$. It is equal to the number of $m \times n$ matrices
    of rank $\le t$ in $\F{q}$. Therefore
    \begin{equation}\label{Eq:VolumeBoule}
    \forall t=0,\ldots,\min(n,m), \quad \mathcal{B}_t = \sum_{i=0}^t {\mathcal{S}_i  }.
    \end{equation}

\end{itemize}

\subsection{Bound on spheres and balls in rank metric}
These quantities (\ref{Eq:VolumeSphere}) and (\ref{Eq:VolumeBoule})
being  not  very convenient to handle, we 
derive the following  bounds, which are not very tight but   sufficient and easy to handle 
for the following of the paper:

\begin{prop} \hfill

For all $t=0,\ldots,\min(n,m),$ we have
\begin{equation}\label{Eq:SphereMetriqueRang}
  \left\{
  \begin{array}{lcl}
  q^{(m+n-2)t - t^2 }  &\le  \mathcal{S}_t \le & q^{(m+n+1)t - t^2 },\\
   q^{(m+n-2)t - t^2 } &\le \mathcal{B}_t  \le & q^{(m+n+1)t - t^2 +1 }.
  \end{array}
  \right.
\end{equation}
\end{prop}

\begin{proof}
It is clearly true for $t=0$. Now, for $t = 1,\ldots,\min(n,m)$, we have 
\begin{itemize}
 \item {\em Volume of spheres:} Formula for
   (\ref{Eq:SphereMetriqueRang}) can be rewritten   
\[
  \mathcal{S}_t  =  q^{(m+n)t - t^2} \prod_{j=0}^{t-1}{ \frac{( 1 - q^{j-n})(1 - q^{j-m})}{1 - q^{j-t}}}.
\]  
For all $j=0,\ldots,t-1$, since $t\le \min(n,m)$ we have
\[
\left\{
\begin{array}{lclcl}
   ( 1 - q^{t-n-1})(1 - q^{t-m-1})  &\le&  ( 1 - q^{j-n})(1 - q^{j-m})&\le&  1, \\
              1                     &\ge&    1 - q^{j-t}  &\ge&  1 - \frac{1}{q}  = \frac{q-1}{q}.  
\end{array}
\right. 
\]
For the same reason $(1 - q^{-1}) \le  (1 -  q^{t-\min(n,m)-1})$. Therefore 
\[
               (1 - 2q^{-1} + q^{-2})    \le  ( 1 - q^{t-n-1})(1 - q^{t-m-1}).
\] 
Since  $q$ has to be the power of a prime, we have $q \ge 2$. This implies that   $(1-2q^{-1}) \ge 0$ and  that $
q^{-2} \le (1 - q^{t-n-1})(1 - q^{t-m-1})$.
Moreover under the same condition we have  $q/(q-1) \le q$. 
Therefore every term of the product satisfies 
\[
\forall j =0,\ldots, t-1, \quad   q^{-2}   \le  \frac{( 1 - q^{j-n})(1 - q^{j-m})}{1 - q^{j-t}} \le q. 
\]
And we obtain the desired result for the volume of the sphere of radius $t$. 

\item {\em Volume of balls}: Since the volume of a ball is larger than that of a sphere with the same radius, the lower bound 
comes directly from the lower bound for spheres. Let $L = m+n+1$,
the upper bound  on the volume of spheres implies that 
\[
\mathcal{B}_t  \le \sum_{i=0}^{t}{ q^{L i - i^2 }} =  
q^{L t - t^2}  \left( 1 + \sum_{i=0}^{t-1}{q^{(L - i-t)(i-t) } } \right).
\] 
For $i$ from $0$ to $t-1$ we have   $ L-i-t \ge L - t$ and $(i - t) < 0$. Therefore 
$q^{(L-i-t)(i-t)} \le q^{(L-t)(i-t)}$. And by changing variables, we finally obtain  
\[
\mathcal{B}_t  \le 
q^{L t - t^2}  \left( 1 + \sum_{j=1}^{t}{q^{-(L -t)j} } \right).  
\] 
The  geometrical sum  can easily computed. Namely, 
\[
 1 + \sum_{j=1}^{t}{q^{-(L - t)j} } = \frac{ 1 - q^{-(L -t)(t+1)}}{1- q^{-(L -t)}} \le  \frac{ 1}{1- q^{-(L -t)}}.
\]
Since $t \le \min(m,n)$ and $q \ge 2$,  we have $1/(1 - q^{-(L -t)} ) \le 2 \le q$. This proves  
 the result for the volume of  $\mathcal{B}_t$.
\end{itemize}
$\blacksquare$
\end{proof}
\medskip

\subsection{Minimum rank distance of a code}

A code $\mathcal{C}$ of length $n$ and of size $M$ over $\F{q^m}$ is a
set of $M$ vectors of length $n$ over  $\F{q^m}$. Its
minimum  rank distance  is defined by 
 \begin{defi} \label{Defi:DistanceRangMinimale} \hfill

Let  $\mathcal{C}$ be a code over  $\F{q^m}$, then 
$ d \stackrel{def}{=} \min_{\Vec{c}_1 \ne  \Vec{c}_2  \in \mathcal{C}
 }(\Rang(\Vec{c}_1 - \Vec{c}_2 )) $
is called {\em minimum rank distance} of  $\mathcal{C}$. 
\end{defi}

If the code is $\F{q}$-linear (it is most often the case when considered as a matricial code) 
or even linear and  since rank metric is invariant by translation, the {\em minimum rank distance} 
of the considered code is 
\begin{equation}\label{Eq:DistMinLin}
d = \min_{\Vec{c} \ne \Vec{0}  \in \mathcal{C} }(\Rang(\Vec{c})).
\end{equation}
If $d$ is the minimum rank distance of $\mathcal{C}$ we will say that 
$\mathcal{C}$ is a $(n,M,d)_r$-code. Moreover if the code is linear of dimension $k$ we will
say that it is a $[n,k,d]_r$-code.

\subsection{Transposed code}

A $(n,M,d)_r$-code over $\F{q^m}$ can be seen as a $(m,M,d)_r$-code over 
$\F{q^n}$ by the following procedure : As in definition \ref{Defi:RangVecteur}, 
any row vector $\Vec{x} \in \F{q^m}^n$  of rank $t$ corresponds to a $q$-ary $m\times n$ 
matrix $X$ by  extending its components row-wise on the chosen basis $\mathcal{B}$ of $\F{q^m}/\F{q}$. 
Consider the following mapping 
\[
\begin{array}{lcl}
 \mathcal{T}:~ \F{q^m}^n & \longrightarrow & \F{q^n}^m \\
\Vec{x} = \left( \sum_{i=1}^{m}{x_{1i} \beta_i},
  \ldots,\sum_{i=1}^{m}{x_{ni} \beta_i } \right) 
&\stackrel{\mathcal{T}}{\longmapsto}& \Vec{x}^T =  \left( \sum_{j=1}^{n}{x_{j1} \gamma_j}, \ldots,\sum_{j=1}^{n}{x_{jm} \gamma_j } \right),
\end{array}
 \]
where $\gamma_1,\ldots,\gamma_n$ is a basis of  $\F{q^n}/\F{q}$.
Clearly  the mapping   $\mathcal{T}$ is a rank-preserving  isomorphism between
$\F{q^m}^n$ and $\F{q^n}^m$.  By using this mapping we define the notion of transposed code.

\begin{defi}[Transposed code -- see \cite{Gabidulin/Pilipchuk:2006}]\label{Defi:CodeTranspose} \hfill

\begin{itemize}
\item Let $\Vec{x} \in \F{q^m}^n$, then $\Vec{x}^T = \mathcal{T}(\Vec{x})$ is called {\em transposed vector} of $\Vec{x}$. 
\item  Let $\mathcal{C}$ be a $(n,M,d)_r$ code over  $\F{q^m}$, the  $(m,M,d)_r$-code over $\F{q^n}$ 
  defined by $\mathcal{C}^T = \{ \Vec{x}^T ~|~\Vec{x} \in \mathcal{C} \}$,  is called {\em transposed code} of $\mathcal{C}$.
\end{itemize}
\end{defi}

The definition and properties of transposed codes lead us to consider only the case where
$n \le m$, the other cases being symmetric by 
taking the corresponding transposed code.

\begin{remark} \hfill 

By properties of $\mathcal{T}$, if a code $\mathcal{C}$ is $\F{q}$-linear then 
its transposed code $\mathcal{C}^T$ is also $\F{q}$-linear. However $\mathcal{T}$ being only $\F{q}$-linear and
not $\F{q^m}$ linear, if $\mathcal{C}$ is linear then   $\mathcal{C}^T$ has no reasons to be linear.
\end{remark}

\section{Upper bounds and perfect codes}
\label{Section:CodesParfaits}

In this section we recall a Singleton-like bound for rank metric codes
and state  an equivalent to the sphere-packing bound. 
We show that there are no perfect codes in rank metric.
 
\begin{theorem}\hfill 

Let  $\mathcal{C}$ be a  $(n,M,d)_r$ code over  $\F{q^m}$. We have
\begin{itemize}
  \item {\em  Singleton-like  bound:}  
    $M \le q^{\min{( m(n-d+1), n(m-d+1)) }}.$  
  \item {\em Sphere packing-like  bound:}  If   $t = \lfloor (d-1)/2 \rfloor$, then 
    \begin{equation}\label{Eq:SpherePacking}
    M  \times  \mathcal{B}_t 
    \le q^{mn},
    \end{equation}
\end{itemize}
\end{theorem}

  For the proof of Singleton-like bound see \cite{Gabidulin:1985,OGHA:2003}. 
The proof of the sphere-packing bound comes from the fact that, for rank metric, two balls
of radius $t = \lfloor (d-1)/2 \rfloor$ centered on codewords do not intersect. Thus, the
full  covering has size less than the whole space. The proof is
identical as for codes in  Hamming metric. 

Now the {\em Sphere-packing bound} raises a question: If we define a 
perfect code for rank metric as a $(n,M,d)_r$-code over $\F{q^m}$ such
that  
  $M  \times \mathcal{B}_t =  q^{mn}$,
does a such code exist and for which set of parameters ?

The following proposition answers the question

\begin{prop}
  There are no perfect codes in rank metric.
\end{prop}

\begin{proof}
  Suppose on the contrary that a  perfect code does exist with
  parameters $(n,M,d)_r$ over  $\F{q^m}$, that is suppose that
  \[
  M \times  \mathcal{B}_t =  q^{mn}.
  \]
 Without loss of
  generality we can assume that $n\le m$ (Else consider the transposed code). 
The right part of the inequality (\ref{Eq:SphereMetriqueRang}) on
  the volume of balls  implies that 
\[
M   q^{(m+n+1)t - t^2 + 1}  \ge q^{mn}. 
\]
Moreover, from Singleton bound we have   $M \le q^{m(n-d+1)}$. Since $t
= \lfloor (d-1)/2 \rfloor$ this implies that $M \le  q^{m(n-2t)}$. 
Therefore  
\[
 q^{(m+n+1)t - t^2 + m(n - 2t) + 1}   \ge   q^{mn}. 
\]
By taking the base $q$ logarithm of the inequality and by reordering
the terms, we obtain   
\[
    (n - m)t  \ge t^2 - t +1 . 
\]
By hypothesis $n-m \le 0$ and $t > 0$. Therefore we must have 
 $t^2 -t - 1 \le 0$. Since $t$ is integer the only possibility is
 $t=1$ and accordingly $n=m$. In that case however the
 formula that parameters have to satisfy is  $M  \times 
 \mathcal{B}_1 =  q^{n^2}$. Hence 
 \[
   \underbrace{q^{n(n-2)}}_{\mbox{\small{Singleton}}} \frac{q^{2n} - 2 q^n + q}{q-1}  \ge    M   \underbrace{\frac{q^{2n} - 2 q^n + 1}{q-1} +1 }_{\mathcal{B}_1} = q^{n^2}, 
\]  
which implies 
\begin{equation} \label{Eq:Qperfect}
1 - \frac{2}{q^n} + \frac{1}{q^{2n-1}} \ge q-1.
\end{equation}
This inequality cannot be satisfied for $q \ge 2$.

\noindent $\blacksquare$
\end{proof}

Though no perfect codes exist, we will see in section \ref{Section:Densite} that considering the 
case where $q=2$ and $m=n$, there exist families
of linear codes that are {\em almost} perfect that is, given any integer $I < 2^{n^2}$, there is
a $(n,M,d)_r$ code such that $I < M \mathcal{B}_{t} <  2^{n^2}$.

\section{A Varshamov--Gilbert like bound}

Until now we have obtained bounds and results on the non-existence of
codes in rank metric with given parameters. What then of the existence
of codes ? We  define an equivalent of the famous 
Varshamov--Gilbert bound for the rank metric.  An alternative definition was given in \cite{Gadouleau/Yan:2006}, but 
here it is more convenient to establish another one for future proofs.
We also define what does mean for a code to be on GV, and give an asymptotic equivalent
of the bound under some conditions on $m$ and $n$.

\begin{prop} \hfill

Let $m,n,M,d$ be positive integers. If
 \begin{equation} \label{Eq:GV}
M  \times \mathcal{B}_{d-1} < q^{mn}, 
\end{equation}
then there exists a  $(n,M+1,d)_r$-code  over $\F{q^m}$.
\end{prop}

\begin{proof}
We suppose given integers $n$ and $d$, $M$ satisfying equation
(\ref{Eq:GV}). We will prove the proposition by constructing a code 
with required parameters. 

\begin{itemize}
  \item {\em Constructing a code with two elements:} we choose
    randomly a vector $\Vec{c} \in \F{q^m}^n$. Since $M \ge 1$, and
    from equation (\ref{Eq:GV}), we have that $  \F{q^m}^n \setminus  \mathcal{B}(\Vec{c},d-1)
    \ne \emptyset$. Therefore  there is at least a vector  $\Vec{d} \in \F{q^m}^n \setminus \mathcal{B}(\Vec{c},d-1)$. The code with two elements formed by  $\Vec{c}$ and $\Vec{d}$ has minimum
    rank distance at least $d$.

  \item {\em Induction step:} Suppose that we have constructed a
    $(n,M',d)$-code $\mathcal{C}$ where $M' \le M$. Let us consider the set formed
    by the union of balls of radius $d-1$ centered on the codewords,
    we have 
    \[
\mathcal{V} \stackrel{def}{=} \bigcup_{\Vec{c} \in \mathcal{C}} \mathcal{B}(\Vec{c},d-1).
\]
This implies in particular that $|\mathcal{V}| \le M \times
\mathcal{B}_{d-1}$. Therefore, from equation  (\ref{Eq:GV})
there exists a vector $\Vec{a} \in \F{q^m}^n \setminus
\mathcal{V}$. Therefore $\mathcal{C}' = \mathcal{C} \cup \{ \Vec{a}
\}$ is a $(n,M'+1,d)_r$-code.
\end{itemize}
$\blacksquare$
\end{proof}

From this we now deduce another definition: 

\begin{defi}\label{Defi:CodeSurGV} \hfill

A $(n,M,d)_r$-code is
 said to be on GV if
 \begin{equation} \label{Eq:GV2}
(M-1)  \times  \mathcal{B}_{d-1} < q^{mn} \le  M \times \mathcal{B}_{d-1}, 
\end{equation}
\end{defi}

We can wonder if as it is the case in Hamming metric we can obtain an
asymptotic value for the {\em minimum rank distance} of a code on GV ? 

\begin{prop}  \hfill
   
Consider a $(n,M,d)_r$-code  $\mathcal{C}$ over  $\F{q^m}$ where $m\ge
n$. Then, if $\mathcal{C}$  is on GV   we have 
\[
\frac{d}{m+n}
\stackrel{n \rightarrow + \infty}{\sim}   ~ \frac{1}{2} -  \frac{ \sqrt{\log_q{M}}}{m+n} 
\sqrt{1 + \frac{(m-n)^2}{4 \log_q{M}}},   
\]
provided $mn \ge \log_q M =  \lambda(n)(m+n)$, where  $\lambda(n) = o(n)$
tends to $+\infty$ with  $n$.
\end{prop}

\begin{proof}
  By taking the base $q$ logarithm of the inequalities
  (\ref{Eq:SphereMetriqueRang}), 
  we obtain from property (\ref{Eq:GV2}) that
  \[
\left\{
  \begin{array}{l}
    mn \le (m+n+1)(d-1) - (d-1)^2 + 1 + \log_q M,  \\
     \log_q(M-1) + (m+n-2)(d-1) - (d-1)^2  < mn. 
  \end{array}
\right.
  \]
Since $M \ge 2$ we have further that 
  $\log_q(M-1) \ge \log_q(M) -  \log_q(2) \ge \log_q(M) - 1$. Hence  the minimum distance of the 
code must satisfy 
  \[
  \left\{
  \begin{array}{l}
    0  \le -d^2 + (m+n+3)d +\log_q M -mn -(m+n +1), \\
    0 \ge  -d^2 + (m+n)d +\log_q M  - mn -(m+n).
  \end{array}
  \right.
  \]
  The inequalities are given by second order equations whose  discriminant are respectively 
  \[
   \begin{array}{l}
     \Delta_1 = (m-n)^2 + 4 \log_q(M) + 2(m+n) +5, \\
     \Delta_2 = (m-n)^2 + 4 \log_q(M) - 4(m+n).
  \end{array}
  \]
  Therefore the minimum distance of a code on GV satisfies the inequalities 
  \[
    \frac{1}{2} - \frac{ - \sqrt{\Delta_1} + 3}{2(m+n)} \le \frac{d}{m+n} \le \frac{1}{2} - \frac{\sqrt{\Delta_2}}{2(m+n)}.
  \]
  Under the conditions of the theorem ($\log_q{M} = \lambda(n)(m+n)$,
 where $\lambda(n) = o(n)$ and  tends to infinity with $n$),
 it is not very difficult to complete the proof of the proposition.

\noindent $\blacksquare$ 
\end{proof}

\begin{example} \hfill

  A special case is when $m=n$ and for a family of constant rate
  codes $ 0 < R <1$ that is 
  \[
  \log_q M = n^2 R.
  \]
  In that case we have 
  \[
  \frac{d}{n} \sim 1 - \sqrt{R}.
  \]
  This result implies that the ratio of the minimum rank distance 
  on the length of the code is asymptotically constant. 
\end{example}

\begin{remark} \hfill

Defining an equivalent of GV bound is less relevant in rank metric than in Hamming metric. 
Namely, in section \ref{Section:GabidulinCodes}, we will see that for any set of parameters  codes satisfying
Singleton equality exist, therefore always exceeding GV bound.  Moreover, they are at least additive and can easily 
be constructed from linear codes and overall we know how to construct a polynomial-time decoder. Therefore this case 
is radically different from the case of Hamming metric  where  there  no MDS-binary codes exist. 
Random binary codes are proved to be on GV, which  gives a benchmark for codes that  can be constructed.

In rank metric the main interest of establishing the GV bound 
is to study how random codes are good regarding this bound - this is the object of next section. 
Another interesting question is to measure the interval between the minimum rank distance of codes 
on GV and Singleton bound. Bounds were given in \cite{Gadouleau/Yan:2006} and simulation were made
on the case where the ratio $n/m$ is constant $\le 1$.
\end{remark}

\section{Random codes}

Random codes provide benchmarks for the codes that can be constructed.
In Hamming metric, random $\F{q}$-linear codes are good. 
They are even often better than families of codes for which we know
 polynomial-time decoders.  In addition it was shown that linear codes
 asymptotically achieve GV-bound. 
The question is the same concerning rank metric. Where are random
codes located compared with GV bound ?

First we define what are random codes

\begin{defi}\hfill

  \begin{itemize}
  \item A random code of length $n$ and of size $M$ over $\F{q^m}$ is a set of $M$ distinct uniformly 
chosen vectors of $\F{q^m}^n$.
  \item A random $\F{q}$-linear code of length $n$ and of size $M=q^{K}$ is defined 
    \[
    \mathcal{C} = \left\{ \sum_{j=1}^K{\lambda_j \Vec{b}_j},\quad  \forall j=1,\ldots,K, ~|~ \lambda_j \in \F{q}\right\}, 
    \]
    where for all $j=1,\ldots,K$,  $\Vec{b}_j \in \F{q^m}^n$ are $\F{q}$-linearly independent uniformly chosen vectors.
\end{itemize}
\end{defi}

In this  section we first establish the probability distribution of 
the minimum rank distance of random codes and
of  random $\F{q}$-linear codes. After upper bounding this
distribution,  we show that the probability that the 
minimum rank distance of the code is outside some interval decreases
exponentially whenever the length of the codes increases. 
This enables us to prove the following theorem, which is the main
result of the section.

\begin{theorem}\label{Theo:CodeAleatoire} \hfill

Let $\mathcal{C}$ be a $(n,M,d)_r$-random ($\F{q}$-linear or not) code
over $\F{q^m}$. Let $N= M-1$, if $\mathcal{C}$ is $\F{q}$-linear and $N = M(M-1)/2$ if not.
Suppose moreover that  $n \le m$ and $q^{\alpha mn} \ge  N \ge 3 \log_q(n) q^{m+n}$ for some
 $\alpha <1$. If 
\[
\mathcal{M}_n \stackrel{def}{=} \frac{m+n}{2} - \sqrt{ \frac{(m-n)^2}{4} + \log_q N},
\]
then  the expectation $E(d)$ and the variance $Var(d)$ of the minimum
rank distance of $\mathcal{C}$  satisfy for $n \rightarrow \infty$
\[
\left\{
\begin{array}{lcl}
E(d)   &=&  \mathcal{M}_n + O(1), \\
Var(d) &=& O(\mathcal{M}_n). 
\end{array}
\right.
\]  
\end{theorem}
Since, in the hypotheses of the theorem the $\mathcal{M}_n$ tends to infinity with $n$,
an immediate corollary of the theorem obtained by applying 
the well-known  Chebyshef's inequality is   
\begin{corollary} \hfill

Let $(\mathcal{C}_n)$ be a family of $(n,M_n,d_n)_r$-random codes over
$\F{q^m}$ where
\begin{itemize}
  \item $m \ge n$, 
  \item $q^{\alpha mn} > M_n > 3\log_q(n) q^{m+n}$ for a fixed $\alpha < 1$.
\end{itemize}
Then let $\psi$ be an increasing function of $n$. We have 
\[
\Pr \left( |d_n - E(d_n)| > \psi(n)\sqrt{n} \right) \stackrel{n \rightarrow \infty}{\longrightarrow} 0  
\]
\end{corollary}
This corollary shows that, whenever the conditions are satisfied,  
the minimum distance of random codes corresponds to its expectation
with a probability as small as possible.

Therefore, in the case of random $\F{q}$-linear  codes that is,
whenever $N=M-1$ we deduce a result analog to the result in Hamming metric: 
\begin{corollary} \hfill

  Under the condition of the previous corollary, random $\F{q}$-linear codes are on GV.

\end{corollary}

\subsection{A bound on the minimum rank distance distributions}

We first deal with random codes and then with random $\F{q}$-linear codes.

\begin{paragraph}{Random codes}
Let $\mathcal{C}$ be a
     $(n,M,d)_r$-random code over $\F{q^m}$. We define the following random
variables:  
\[
\forall i=1, \ldots,n, \quad 
\left\{
\begin{array}{l}
\mathcal{A}_i = \left| \left\{ (\Vec{c}_1,\Vec{c}_2) \in \mathcal{C}^2 ~|~
\Rang(\Vec{c}_1 - \Vec{c}_2) = i \right\} \right|, \\
\mathcal{D}_i  =  \cup_{t=1}^i \mathcal{A}_i =  
|   \left\{ (\Vec{c}_1,\Vec{c}_2) \in \mathcal{C}^2, \Vec{c}_1 \ne \Vec{c}_2 ~|~
\Rang(\Vec{c}_1 - \Vec{c}_2) \le i \right\} |.
\end{array}
\right.
\]
The random variable  $\mathcal{A}_i$ counts the number of pairs of codewords at
distance exactly equal to $i$ whereas the random variable $\mathcal{D}_i$ counts the
number of pairs of distinct codewords at distance  $\le i$. From this, 
the random variable $d$ specifying the minimum rank
distance of $\mathcal{C}$ satisfies the distribution  
\[
\forall i=1,\ldots,n, \quad p_i \stackrel{def}{=}  \Pr(d=i) = \Pr( \mathcal{D}_{i-1} =0,~ \mathcal{D}_i \ge 1).
\] 
From Bayes' rule we have
$
\Pr(d = i ) = \Pr( \mathcal{D}_{i} \ge 1 ~|~ \mathcal{D}_{i-1} = 0)
\Pr(\mathcal{D}_{i-1} = 0)$.
For all $i=1,\ldots,n$, the random variable  $\mathcal{D}_i$ can be
expressed as
\[
\mathcal{D}_{i} = \sum_{\Vec{c} \ne \Vec{d} \in
  \mathcal{C}^2}{\mathbf{1}_{\small \Rang(\Vec{c}- \Vec{d}) \le i }}
\]
Therefore $\mathcal{D}_i$ is equal  to $0$ if and only if for all $\Vec{c} \in \mathcal{C}$
and $\Vec{d} \in \mathcal{C}$  such that $\Vec{c} \ne \Vec{d}$, $\Rang(\Vec{c}- \Vec{d}) \ge i + 1$. 
Since the cardinality of the code is equal to $M$ and by taking into account the symmetries ({\em i.e.} the fact that
$ \Rang(\Vec{c}- \Vec{d}) = \Rang(\Vec{d}- \Vec{c})$), we have

\begin{equation}\label{Eq:ProbBoule1}
\Pr(\mathcal{D}_{i-1} = 0) = \left( 1 -  \frac{\mathcal{B}_{i-1}}{q^{mn}} \right)^{M(M-1)/2}.
\end{equation}
We also have 
\[
\Pr( \mathcal{D}_{i} \ge 1 ~|~ \mathcal{D}_{i-1} = 0) = 1 - \Pr(
\mathcal{D}_{i} = 0  ~|~ \mathcal{D}_{i-1} = 0). 
\]
It is clear equally that, since $\mathcal{D}_i = \mathcal{A}_i \cup \mathcal{D}_{i-1}$, we have 
\[
\Pr(\mathcal{D}_i = 0 ~|~  \mathcal{D}_{i-1} = 0) =   \Pr(\mathcal{A}_i = 0 ~|~  \mathcal{D}_{i-1} = 0).
\]
Since the words of the codes are randomly chosen one can see that 
\[
 \Pr(\mathcal{A}_i = 0 ~|~  \mathcal{D}_{i-1} = 0) =  \left( 1 -
  \frac{\mathcal{S}_i}{q^{mn} - \mathcal{B}_{i-1}} \right)^{M(M-1)/2}
\]
Hence we finally obtain 
\begin{equation}\label{Eq:ProbBoule2}
\Pr( \mathcal{D}_{i} \ge 1 ~|~ \mathcal{D}_{i-1} = 0) = 1 - \left( 1 -
  \frac{\mathcal{S}_i}{q^{mn} - \mathcal{B}_{i-1}} \right)^{M(M-1)/2}. 
\end{equation}
\end{paragraph}

\begin{paragraph}{Random $\F{q}$-linear codes}

The approach for $\F{q}$-linear codes is the same except that $\F{q}$-linearity modifies the definition
for the minimum rank distance of the code, see formula (\ref{Eq:DistMinLin}). If we set  
\[
\forall i=1, \ldots,n, \quad 
\left\{
\begin{array}{l}
\mathcal{A}_i = \left| \left\{ \Vec{c} \in \mathcal{C} ~|~
\Rang(\Vec{c}) = i \right\} \right|, \\
\mathcal{D}_i  =  \cup_{t=1}^i \mathcal{A}_i =  
|   \left\{ \Vec{c} \in \mathcal{C}\setminus \{\Vec{0} \}, ~|~
\Rang(\Vec{c}) \le i \right\} |,
\end{array}
\right.
\]
we have 
\[
\forall i=1,\ldots,n, \quad p_i \stackrel{def}{=}  \Pr(d=i) = \Pr( \mathcal{D}_{i-1} =0,~ \mathcal{D}_i \ge 1).
\] 
Since 
\[
\mathcal{D}_i = \sum_{\Vec{c} \in \mathcal{C}\setminus \{\Vec{0}\}}{\Vec{1}_{\Rang(\Vec{c}) \le i}},
\]
and since  the vectors $\Vec{b}_j$ forming a $\F{q}$-basis for $\mathcal{C}$ are uniformly chosen, 
one can check that $\Pr(\Rang(\Vec{c}) = i ~|~ \Vec{c} \in \mathcal{C}) = 
\Pr(\Rang(\Vec{c}) = i) =  \mathcal{B}_i/ q^{mn}$. Therefore  

\begin{equation}\label{Eq:ProbLineBoule1}
\Pr(\mathcal{D}_{i-1} = 0) = \Pr(\forall \Vec{c} \in \mathcal{C}\setminus \{\Vec{0} \}, \Rang(\Vec{c}) \ge i) = \left( 1 -  \frac{\mathcal{B}_{i-1}}{q^{mn}} \right)^{M-1}
\end{equation}
and 
\begin{equation}\label{Eq:ProbLineBoule2}
\Pr( \mathcal{D}_{i} \ge 1 ~|~ \mathcal{D}_{i-1} = 0) = 1 - \left( 1 -
  \frac{\mathcal{S}_i}{q^{mn} - \mathcal{B}_{i-1}} \right)^{M-1}
\end{equation}
\end{paragraph}

From the previous paragraphs,  
 by multiplying (\ref{Eq:ProbBoule1}) by 
(\ref{Eq:ProbBoule2}) in the general case or (\ref{Eq:ProbLineBoule1}) by (\ref{Eq:ProbLineBoule2}) 
in the $\F{q}$-linear case 
and since $\mathcal{B}_i = \mathcal{B}_{i-1} +
\mathcal{S}_i$ one obtains the following proposition

\begin{prop}
\hfill 

  Let $\mathcal{C}$ be a $(n,M,d)_r$ random code over $\F{q^m}$. 
  Let $N= M-1$, if $\mathcal{C}$ is $\F{q}$-linear and $N = M(M-1)/2$ if not.
 Let
  \[
  \forall i=1,\ldots, n, ~ p_i = \Pr(d = i).
  \]
  Then we have
  \begin{equation} \label{Eq:ProbaDist}
  \forall i=1, \ldots, n, \quad  p_i = \left(1 -
    \frac{\mathcal{B}_{i-1}}{q^{mn}} \right)^{N} - 
\left( 1 -
    \frac{\mathcal{B}_{i}}{q^{mn}}  \right)^{N},
  \end{equation}
  where $\mathcal{B}_i$ designs the volume of the ball of radius
  $i$ in $\F{q^m}^n$.
\end{prop}

From formula (\ref{Eq:ProbaDist}) and using the fact that for any positive integer $N$, 
\[
\forall a \ge b \ge 0,~ a^N - b^N = (a - b)
\sum_{j=0}^{N-1}{a^j b^{N-1-j}} \le N (a-b) \left( \max(a,b) \right) ^{N-1},
\]
we deduce the following upper bound
\begin{equation}\label{Eq:BorneProbaDistance}
   p_i \le  \frac{N \mathcal{S}_i}{q^{mn}} \left( 1 -
    \frac{\mathcal{B}_{i-1}}{q^{mn}} \right)^{N-1}.
\end{equation}

The following lemma will help to prove theorem
\ref{Theo:CodeAleatoire}, by giving ranges for  integer $1 \le i \le n$  where the probability $p_i$
is upper-bounded by an exponential term.

\begin{lemma} \hfill \label{Lemma:BorneDistance}
  
Let $\mathcal{C}$ be a $(n,M,d)_r$ random code over $\F{q^m}$. 
  Let $N= M-1$, if $\mathcal{C}$ is $\F{q}$-linear and $N = M(M-1)/2$ if not. Let
  \[
  \forall i=1,\ldots, n, ~ p_i = \Pr(d = i), \mbox{ then}
  \]
 \begin{itemize}
 \item If $1 \le i \le \frac{m+n}{2} - \sqrt{ \frac{(m-n)^2}{4} + \log_q N + \frac{m+n}{2} +
     \lambda(n)}$, then there is a positive constant $0 < C_1$ such that
   \begin{equation}\label{Eq:BorneIPetit}
   p_i \le C_1 q^{-\lambda(n)}
   \end{equation}
 \item If $N \ge  q^{m+n} \mu(n)$ and if 
$n \ge  i \ge \frac{m+n}{2} - \sqrt{ \frac{(m-n)^2}{4} +  \log_qN  -
  (m+n)  - \log_q\mu(n)}$, then there is a positive constant $C_2$
such that   
\begin{equation} \label{Eq:BorneIGrand}
  p_i \le e^{- C_2 \mu(n)}.
\end{equation}
 \end{itemize}
\end{lemma}

\begin{proof} \hfill

To prove this lemma, we upper-bound  $p_i$ by upper-bounding the inequality  (\ref{Eq:BorneProbaDistance})  
\begin{itemize}
\item The upper bound (\ref{Eq:SphereMetriqueRang}), gives
\[
 p_i \le   \frac{N \mathcal{S}_i}{q^{mn}} \underbrace{\left( 1 -
    \frac{\mathcal{B}_{i-1}}{q^{mn}} \right)^{N-1}}_{< 1} \le  \frac{N \mathcal{S}_i}{q^{mn}} \le q^{(m+n+1)i - i^2 - m n +
 \log_q N}.
\]
The right part of the inequality is smaller than $q^{- \lambda(n)}$ if and
only if $-i^2 + (m+n+1)i    - mn + \log_qN   + \lambda(n) \le
0$. The discriminant of this second order inequation is
\[
\Delta_1 = (m-n)^2  + 1 + 2(m+n) + 4 \log_qN  + 4 \lambda(n).
\]
Therefore if $i$ is smaller than the smallest root of the inequation,
that is if 
\[
i \le \frac{m+n+1}{2} - \sqrt{ \frac{(m-n)^2}{4} + \log_qN + \frac{n+m}{2} + \lambda(n)  + 1/4},
\]
then by  taking $C_1 = q^{-1/4}$ the inequality (\ref{Eq:BorneIPetit}) is satisfied.
\item For the other bound we have
\[
p_i \le  \underbrace{\frac{\mathcal{S}_i}{q^{mn}}}_{< 1} N \left( 1 -
    \frac{\mathcal{B}_{i-1}}{q^{mn}} \right)^{N-1} = e^{(N-1)  \ln \left( 1 -
    \frac{\mathcal{B}_{i-1}}{q^{mn}} \right) + \ln N}.
\]
Since by property of the logarithm function $\forall  0 \le x<1 \ln(1-x) \le -x$, we have 
\[
N \left( 1 -
    \frac{\mathcal{B}_{i-1}}{q^{mn}} \right)^{N-1} \le e^{- (N-1)  
    \frac{\mathcal{B}_{i-1}}{q^{mn}} + \ln N}
\]
Moreover, if we use the  lower bound given in  (\ref{Eq:SphereMetriqueRang}), we have
\[
- (N-1) \frac{\mathcal{B}_{i-1}}{q^{mn}} \le  - q^{(m+n-2)(i-1) - (i-1)^2 - mn + \log_q (N-1)}.
\]
Hence 
\[
N \left( 1 -
    \frac{\mathcal{B}_{i-1}}{q^{mn}} \right)^{N} \le e^{-\mu(n)},
\]
provided in particular that 
$q^{(m+n-2)(i-1) - (i-1)^2 - mn + \log_q (N-1)} \ge \mu(n)$. This
 corresponds to  
\[
- i^2 +(m+n)i  - (m+n) + 1  -mn +  \log_q (N-1) - \log_q \mu(n) \ge 0.
\]
The   discriminant  of the second order equation is equal to   
\[ 
\Delta_2 =  (m-n)^2  +1 + 4 (\log_q (N-1) - (m+n) - \log_q \mu(n)). 
\]
Since by hypothesis we have  $N \ge q^{m+n}\mu(n)$ we can show that
$\Delta_2 \ge 0$ and  since there exists a constant $C$ such that 
$\forall N > 1 \log_q(N-1) \le \log_qN + C$,  for all integer $i$ such
that    
\[
n \ge i \ge \frac{m+n}{2} - \sqrt{ \frac{(m-n)^2}{4}  +   \log_q N - (m+n) -
  \log_q \mu(n)},
\]
we also obtain $ p_i \le e^{- C_2 \mu(n)}$, where $C_2 = q^C$ is a constant
\end{itemize}

$\blacksquare$
\end{proof}

The asymptotic behavior of the upper bounds on the $p_i$  depend on how the
parameters $m$ and $N$ increase when the length $n$ of the code
increases. The idea to prove theorem \ref{Theo:CodeAleatoire} is to 
 consider positive increasing functions $\lambda$
 and $\mu$, sufficiently fast but not too fast so that the
 contribution to that the probability that the minimum rank distance
 of the code be equal to $i$ outside the interval given by
\[
\small
   \frac{m+n}{2} - \sqrt{ \frac{(m-n)^2}{4} + \log_qN +
   \frac{n+m}{2} + \lambda(n)} < i <  \frac{m+n}{2} - \sqrt{ \frac{(m-n)^2}{4} +  \log_qN  - (m+n)  - \log_q\mu(n)}
\]
tends asymptotically to $0$.

\subsection{Proof of the theorem}

Now we will prove theorem \ref{Theo:CodeAleatoire}.

\begin{paragraph}{Proof of theorem \ref{Theo:CodeAleatoire}} \hfill 

By definition the expectation of the minimum rank distance is given by
\[
E(d) = \sum_{i=1}^n{i p_i}.
\]
 We shall use the
results obtained in lemma \ref{Lemma:BorneDistance} with
 $\lambda(n) = \mu(n) = 3 \log_{q}n $. This choice is arbitrary but 
sufficient for the rest of the proof.

The interest is that, since $\log_q N$
increases faster than $(m+n)$ 
 the contribution of  functions $\lambda$ and $\mu$  under the radical
 is always negligible 
compared to the contribution of the other functions. We define  
\[
\left\{
\begin{array}{l}
a_n = \left\lfloor \frac{m+n}{2} - \sqrt{ \frac{(m-n)^2}{4} + \log_q N + \frac{m+n}{2} +
     \lambda(n)} \right\rfloor,\\
   b_n = \left\lfloor  \frac{m+n}{2} - \sqrt{ \frac{(m-n)^2}{4} +  \log_qN
     - (m+n)  - \log_q\mu(n)} \right\rfloor +1 .
\end{array}
\right.
\]
From the hypotheses, and the choices $\lambda$ and $\mu$, for
sufficiently large $n$,  $a_n > 1$ and $b_n < n $. From the results of 
lemma \ref{Lemma:BorneDistance}, for all $1 \le i \le a_n$ and
$b_n \le i \le n$, there is a constant $C$ such that 
$p_i \le C / n^{3}$. Therefore 
\[
\left\{
\begin{array}{lclcl}
  \sum_{i=1}^{a_n}{i p_i} &\le&  \frac{C}{n^3}  \sum_{i=1}^{n}{i}   &=&  O\left(\frac{1}{n} \right),\\
  \sum_{i=b_n}^{n}{i p_i} &\le&   \frac{C}{n^3} \sum_{i=1}^{n}{i}   &=&  O\left(\frac{1}{n} \right).
\end{array}
\right.
\]
Moreover since we deal only with positive terms, we have
\[
 a_n  \sum_{i=a_n +1}^{b_n-1}{p_i}    \le \sum_{i=a_n
   +1}^{b_n-1}{i p_i}  
\le b_n \sum_{i=a_n +1}^{b_n-1}{p_i}.
\]
By using once more the results of previous lemma and the fact
that $\sum_{i=1}^n{p_i} =1$ we have  
\[
      1 -   \frac{C}{n^{2}}  \le \sum_{i=a_n +1}^{b_n-1}{p_i} \le 1.
\]
Hence, if we write $\mathcal{M}_n =  \frac{m+n}{2} - \sqrt{\frac{(m-n)^2}{4} + \log_qN}$, then 
\[
     a_n - \mathcal{M}_n -   \frac{C a_n}{n^{2}} +  O\left( \frac{1}{n} \right)
     \le  E(d) -\mathcal{M}_n \le b_n -\mathcal{M}_n + O\left( \frac{1}{n} \right).
\]
We have 
\[
\left\{
\begin{array}{lcl}
  -1 + \sqrt{ \frac{(m-n)^2}{4} + \log_q N}   - \sqrt{ \frac{(m-n)^2}{4} + \log_q N + \frac{m+n}{2} +
     \lambda(n)} &~\le~& a_n - \mathcal{M}_n,\\
   1 +  \sqrt{ \frac{(m-n)^2}{4} + \log_q N} - \sqrt{ \frac{(m-n)^2}{4} +  \log_qN
     - (m+n)  - \log_q\mu(n)} &~\ge~& b_n - \mathcal{M}_n.
\end{array}
\right.
\]
For the integer  $N$ satisfying the hypotheses of the theorem, 
in both cases  the difference between radicals has finite limit 
when $n$ tends to infinity. Therefore the difference between radicals
is equal to $O(1)$

One can equally check that the
quantity $a_n$ does not increase faster than $n^2$  for  $m \ge
n$. 
This implies that   $ \frac{C a_n}{n^2} = O(1)$. 
Therefore 
\begin{equation}\label{Eq:Esperance}
E(d) = \mathcal{M}_n + O(1).
\end{equation}
The variance is given by $Var(d) = E(d^2) -E(d)^2$. From
(\ref{Eq:Esperance}) we have 
\[
E(d)^2 = \mathcal{M}_n^2 + O(\mathcal{M}_n),
\] 
and from the definition of $E(d^2)$.
\[
E(d^2) = \underbrace{\sum_{i= 1}^{a_n}{i^2 p_i}}_{=
  O(1) } +
\underbrace{\sum_{i= b_n}^{n}{i^2 p_i}}_{ = O(1)   } +
\sum_{i=a_n+1}^{b_n-1}{i^2 p_i}.
\]

Using the same approach as before, under the conditions of the
theorem, $\mathcal{M}_n$ tends to infinity with $n$. Therefore $O(1)$ is
negligible compared to $O(\mathcal{M}_n)$. 
\[
  a_n^2 - \mathcal{M}_n^2  + O(\mathcal{M}_n)  \le Var(d) \le  b_n^2 - \mathcal{M}_n^2  + O(\mathcal{M}_n).
\]
Moreover $ a_n^2 - \mathcal{M}_n^2  = (a_n + \mathcal{M}_n)O(1) = O(\mathcal{M}_n)$ and $ b_n^2 - \mathcal{M}_n^2  =
(b_n + \mathcal{M}_n)O(1) = O(\mathcal{M}_n)$, which gives the desired result.

\noindent $\blacksquare$ 
\end{paragraph}

\section{Maximum Rank Distance codes}

Singleton inequality provides an upper bound on the cardinality of codes with 
given parameters. We call optimal codes or MRD ({\em Maximal Rank
  Distance}) codes,  codes satisfying the Singleton equality

\begin{defi}[MRD-codes -- \cite{Gabidulin:1985}] \hfill

A $(n,M,d)_r$-code over $\F{q^m}$ is called MRD if 
\begin{itemize}
  \item $M = q^{m(n-d+1)}$, if $n \le m$.
  \item $M = q^{m(n-d+1)}$, if $n>m$
\end{itemize}
\end{defi}
From this definition it follows that, whenever a code is MRD,  the
corresponding  transposed code is also MRD. 

\subsection{Rank weight distribution of MRD-codes}

In Hamming metric, the weight distribution of MDS-codes is well-known \cite{MacWilliams/Sloane:1977}. 
Gabidulin showed the rank distribution of codes in rank metric could
be expressed by the following formula.
\begin{prop}[\cite{Gabidulin:1985}] \hfill

Let $A_{s}(n,d)$ be the number of codewords of rank $s$ of  a MRD-code
over  $\F{q^m}$. Then  
\begin{equation} \label{Ea:DistriRang}
  A_{d+\ell}(n,d) =  \left[  \begin{array}{c}
n \\
 d + \ell
\end{array} \right]_q
        \sum_{t=0}^{\ell}{(-1)^{t+\ell}  
 \left[  \begin{array}{c}
d + \ell \\
 \ell+t
\end{array} \right]_q
q^{ \ell-t \choose 2} \left(q^{m(t+1)} -1 \right)},
\end{equation}
where $\left[  \begin{array}{c}
n \\
 i
\end{array} \right]_q$ is the  Gaussian binomial.
\end{prop}

Our contribution to this section comes from  
the  simulations we made to evaluate the {\em randomness degree} of
MRD-codes. 
Through  these simulations we obtained that the rank distribution of
 random $\F{q}$-linear codes in  rank metric was almost identical to the weight
distribution of linear MRD-codes. Results are presented in table
\ref{Table:DistriRang}. The table  gives the base $2$ logarithm of the
proportion $ A_{i}(32)/2^{mn}$ for $m \ge 32$. The left-most curve
corresponds to $m=32$, the right-most to $m=40$.
We made simulations for random $\F{q}$-linear codes as well as for 
MRD-codes {\em sufficiently} large with the same parameters. For ranks
significantly greater than the minimum rank distance both curves
coincide very accurately.

\begin{table}[h] \label{Table:DistriRang}
\begin{center}
\resizebox{14cm}{8cm}
{\includegraphics[0cm,0cm][20cm,25cm]{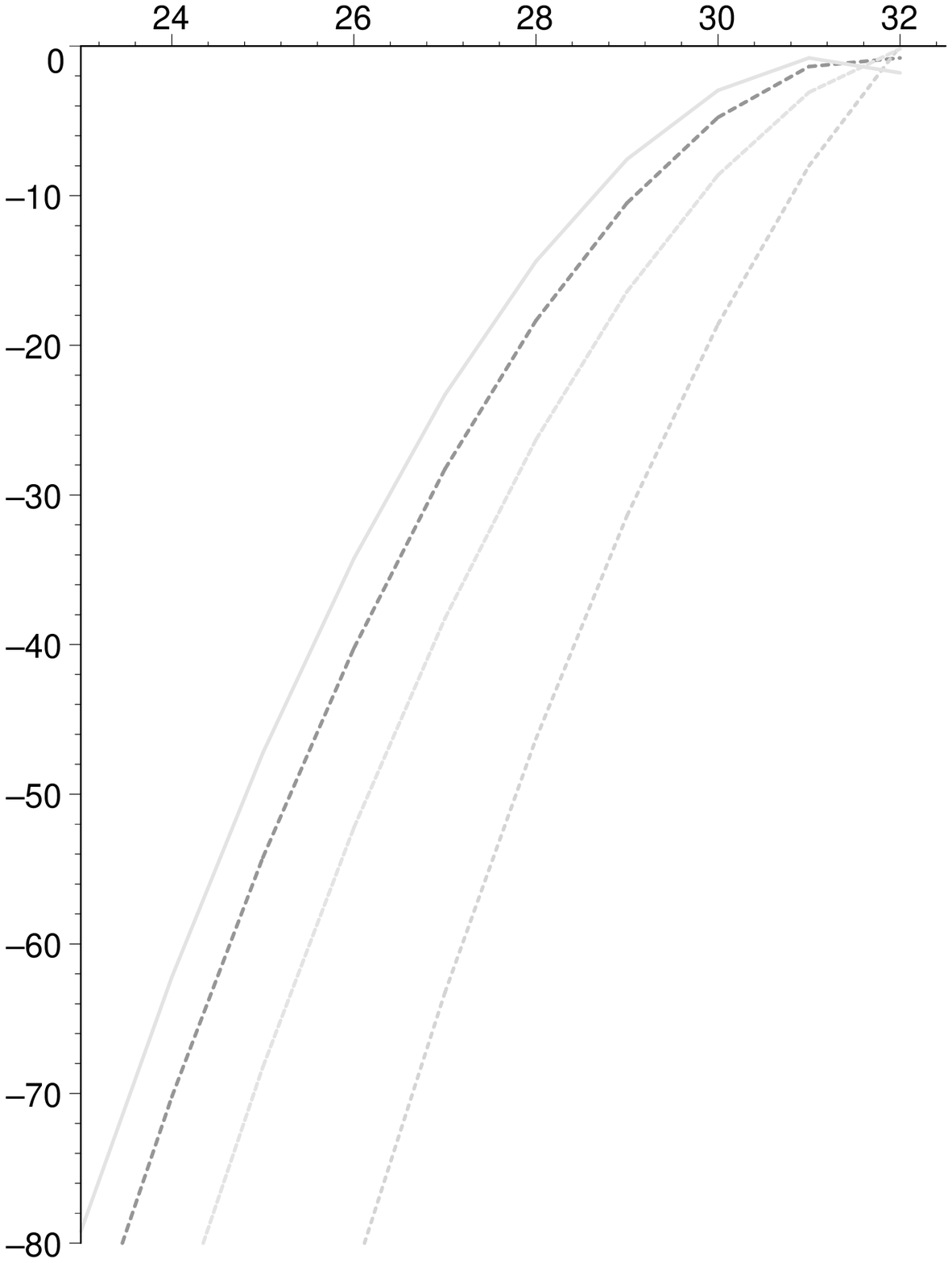}}
\end{center}
\caption{Base 2 logarithm of proportion of words of given rank in an
  MRD-codes  of length $n=32$ over $\F{2^m}$, where    $m =  32,~33,~35,~40$.}
\end{table}

\subsection{Covering density of MRD codes}
\label{Section:Densite}

In section \ref{Section:CodesParfaits} we proved  that no perfect
codes existed in rank metric. 
However a natural question could be: what is the {\em defect} of {\em
  perfectitude} of MRD-codes that is, 
given a $(n,M,d)_r$ MRD-code what is the volume of the space covered
by balls of radius $\lfloor (d-1)/2 \rfloor$ compared to the volume
of the whole space.  The {\em covering density} of the code is thus defined by 
\[
D = \frac{M \mathcal{B}_t }{q^{mn}},
\]
where $t = \lfloor (d-1)/2 \rfloor$.
By using the bounds on the volume of the balls given in  (\ref{Eq:SphereMetriqueRang}), we get 
immediately the following proposition
\begin{prop}[Density of MRD-codes] \label{Prop:Densite} \hfill

Let  $\mathcal{C}$ be a  MRD-code, $(n,q^{m(n-2t)},2t+1)_r$ over $\F{q^m}$.
  The covering density of $\mathcal{C}$ satisfies  
  \[
  \frac{1}{q^{(m-n+2)t + t^2 }}    \le D \le  \frac{1}{q^{(m-n-1)t + t^2 }},
  \]
\end{prop}
The proposition shows that whenever  the length of the code equals the
extension degree, {\em i.e.} $n=m$, which can be the case  for
square-matrix codes, and  
if $n$ tends to $\infty$, then its  covering density is lower bounded
by the quantity  $q^{-t^2 - 2t}$ depending only on the rank error-correcting capability of the code.

\begin{paragraph}{Particular case of rank $1$ correcting MRD codes}
For $1$-error correcting codes where $m=n$, we can express the exact formulas for balls and obtain.
\begin{prop} \label{Prop:CodeQP} \hfill

 A $(n,q^{n-2},3)_r$ MRD-code  over $\F{q^n}$ has a covering 
density equal to  
\begin{equation}
D = \frac{1 - 2 q^{-n} + q^{-2n+1}}{q-1}.
\end{equation}
\end{prop}

There is a special interest in the binary case. In section
\ref{Section:CodesParfaits}, we have proved that there are no perfect
codes in rank metric. We deduce the following corollary from previous proposition

\begin{corollary} \label{Cor:QuasiPerfectCodes} \hfill

Let  $\mathcal{F} =  \{ \mathcal{C}_i \}_{i \ge 2 } $ be a family of  $(i,2^{i-2},3)_r$ MRD-codes over
$\F{2^i}$. If $D_i$ is the covering density of code $\mathcal{C}_i$
then 
\[
\lim_{i \rightarrow \infty}{D_i} = 1.
\] 
\end{corollary}
This means that  $\mathcal{F}$ is a sequence of codes with increasing
length and alphabet that are asymptotically perfect. We say that the
codes in $\mathcal{F}$ are {\em quasi-perfect}.
 
\end{paragraph}

\subsection{A family of decodable optimal codes:  Gabidulin codes}
\label{Section:GabidulinCodes}

In previous section, we established results on MRD-codes before
 knowing if codes with such optimal properties existed. The answer is
 yes and a family of linear MRD-codes was
constructed by  E.~M.~Gabidulin in 1985, see \cite{Gabidulin:1985}. 
In 1991 R.~M.~Roth designed also  a family of  MRD-codes for rank metric in the 
form of $m\times n$-matrix code.  It appeared that both constructions
 were equivalent.  Codes defined by R.~M.~Roth can be obtained by
 considering the matricial  form over the base field of the words of
 some Gabidulin codes \cite{Roth:1991}.

\begin{defi}[Gabidulin code]  \hfill

  Let $\Vec{g} = \Ln{g}$ be a vector of elements of $\F{q^m}$ that are linearly independent
over $\F{q}$, let 

\begin{equation} \label{Eq:MatGeneratrice}
\Vec{G} = \left(
\begin{array}{ccc}
g_1 & \cdots & g_n \\
\vdots & \ddots & \vdots \\
g_1^{[k-1]} & \cdots & g_n^{[k-1]}
\end{array}
\right),
\end{equation}
where $[i]\stackrel{def}{=} q^i$. The code generated by $\Vec{G}$
is the  $k$-dimensional Gabidulin code $Gab_{k}(\Vec{g})$.
$\Vec{g}$ is called  generating 
vector $\Vec{g}$ of the code.
\end{defi}
Gabidulin codes are evaluation codes of so-called linearized  polynomials also called $q$-polynomials or still
{\/\O}re polynomials since they were first introduced by O.~{\/\O}re in the 30's, see 
\cite{Ore:1933,Ore:1934}. Gabidulin codes share many similarities with the well-known 
family of Reed-Solomon codes. In particular the  polynomial-time decoding algorithms are based on some 
Extended Euclidian algorithm for linearized polynomials or a Welch-Berlekamp approach. 
 The
interested reader can refer to the historical genesis of the
algorithms  \cite{Gabidulin:1985,Gabidulin:1991,Roth:1991,Richter/Plass:2004_1,Gabidulin:1991,Loidreau:2005}.

From previous section we know their rank distribution, as well as their covering
density. Since Gabidulin codes are MRD-codes, we can show that 

\begin{prop} \hfill

For all $i \ge 2$,   $Gab_{i-3}(\Vec{\Vec{g}_i})$, 
is a {\em quasi-perfect} code provided  the vectors $\Vec{g}_i \in \F{2^i}^i$ consist 
of components that are linearly independent over $\F{2}$.  
\end{prop}

\begin{proof}\hfill

  Consider the family $\mathcal{F} = \{ Gab_{i-3}(\Vec{g}_i) \}_{i \ge 2}$, where
  $\Vec{g}_i \in \F{2^i}^i$ is formed with linearly independent components. The family $\mathcal{F}$
satisfies the conditions of corollary \ref{Cor:QuasiPerfectCodes}.  

\noindent $\blacksquare$
\end{proof}

\section{Conclusion}

In this paper we presented general results about rank metric and the existence and the properties of
codes in this metric. Problems and properties of 
a such metric  have not been as intensively explored as properties of
Hamming metric. 
Some interesting problems already solved for Hamming metric remain unsolved yet.  
For example, we do not know any equivalent of the MacWilliams
transforms,  which   link  the  minimum rank distance of a linear code
and the minimum rank distance  of its dual. 
Some steps towards this achievement 
can be  found in a work published by  Delsarte
\cite{Delsarte:1978}. In that paper,  matrices with coefficients in a
finite field  are considered as matrices  of quadratic forms and
notion of orthogonality  defined is related to the trace of a product of 
matrices. By this approach, the author   established  an equivalent of
MacWilliams transforms. However, there are no 
simple transformations linking the standard  scalar product 
over finite fields and the trace of the product  of corresponding matrices.

Another interesting point to investigate could be  to construct codes
 with a polynomial-time decoding algorithm 
which are not designed on the basis of the codes introduced in
 \cite{Gabidulin:1985,Roth:1991}. 
Indeed, even subspace subcodes or subfield subcodes 
of  these codes are isomorphic to codes from the same 
family with smaller parameters, see \cite{Gabidulin/Loidreau:2005}. 
Projecting the codes does not affect
their structure conversely to the case of Hamming metric where many
 interesting and widely used codes are subfield subcodes of
 Reed-Solomon codes.

\end{document}